\documentclass[aps,twocolumn,showpacs,superscriptaddress,amsmath]{revtex4}
\usepackage{graphicx}
\newcommand{\bea}{\begin{eqnarray}}
\newcommand{\beq}{\begin{equation}}
\newcommand{\eea}{\end{eqnarray}}
\newcommand{\eeq}{\end{equation}}
\begin{document}
\title{Analysis of nucleus-nucleus collisions at high energies and
Random Matrix Theory}
\vspace{0.5cm}
\author{R.\ G.\ Nazmitdinov}
\affiliation{Departament de F{\'\i}sica,
Universitat de les Illes Balears, E-07122 Palma de Mallorca, Spain}
\affiliation{Bogoliubov Laboratory of Theoretical Physics,
Joint Institute for Nuclear Research, 141980 Dubna, Russia}
\author{E.\ I.\ Shahaliev}
\affiliation{High Energy Physics Laboratory, Joint Institute for
Nuclear Research, 141980, Dubna, Russia} \affiliation{Institute of
Radiation Problems, 370143, Baku, Azerbaijan}
\author{M.\ K.\ Suleymanov}
\affiliation{High Energy Physics Laboratory, Joint Institute for
Nuclear Research, 141980, Dubna, Russia}
\author{S.\ Tomsovic \footnote{permanent address: Department of Physics
and Astronomy, Washington State University, Pullman, WA  99164-2814, USA}}
\affiliation{Max-Planck-Institut f\"ur Physik komplexer Systeme, N\"othnitzer
Stra$\beta$e 38, D-01187 Dresden, Germany}

\date{\today}

\begin{abstract}
We propose a novel statistical approach to the analysis of experimental data obtained
in nucleus-nucleus collisions at high energies which borrows from methods developed
within the context of Random Matrix Theory.  It is applied to the detection of
correlations in a system of secondary particles.  We find good agreement
between the results obtained in this way and a standard analysis based on the
method of effective mass spectra and two-pair correlation function often used in
high energy physics. The method introduced here is free from unwanted background
contributions.
\end{abstract}
\maketitle

\section{Introduction}
There is currently an enormous effort underway to detect signals of possible
transitions between different phases of strongly interacting matter produced
in high energy nucleus-nucleus collisions (cf \cite{1,2}).  It is expected that
in central collisions, at energies that are and will be soon available
at CERN Super Proton Synchrotron (SPS),
BNL Relativistic Heavy Ion Collider (RHIC), and
CERN Large Hadron Collider (LHC), the nuclear density may exceed
by a few orders of magnitude the density of stable nuclei.
At such extreme conditions one would expect that a final product of heavy
ion collisions could present a composite system that consists of
free hadrons, quarks and gluons.
However, quarks and gluons cannot be unambiguously identified
with present detectors. In addition,
the identification, for example, of a quark-gluon plasma is hindered
due to a multiplicity of secondary particles created at these collisions.
In fact, there are numerous additional mechanisms of
particle creation that mask the presence of a quark-gluon plasma.
A natural question arises as to how to identify a useful signal that would
be unambiguously associated with a certain physical process?

Amongst the most popular traditional methods of analysing data produced at
high-energy nucleus-nucleus collisions are: i) the correlation
analysis~\cite{drem}, ii) the analysis of  missing masses~\cite{b2}
and effective mass spectra~\cite{b4}, iii) the interference method of
identical particles~\cite{pod} etc.  It is currently believed that by measuring
event-by-event fluctuations it would be possible to observe anomalies from the onset of
deconfinement \cite{mr}.  Often though, results based on such
methods are sensitive to assumptions made concerning the background measurements
and mechanisms included in the corresponding model.
For example, one of the popular approaches to studying particle production in
heavy ion collisions is through statistical mechanics techniques (cf \cite{a,b,c}).
However, recent theoretical analyses of the multiplicity fluctuations demonstrate that
corresponding results are different in different ensembles and it is sensitive
to conservation laws obeyed by a statistical system \cite{beg}.

Transport theory has played an important role in the interpretation of experimental
results in heavy ion reactions at various energy domains
(see, for example, Refs.\onlinecite{bass,bord}). The underlying concept of
transport models is based on numerical solution of the Boltzmann equation
(the Vlasov-Uehling-Uhlenbeck, the Boltzmann-Uehling-Uhlenbeck (BUU) {\it etc}).
However, exact numerical solutions, for example,  of the BUU equation
are very difficult to obtain and, therefore, various approximate treatments have been
developed \cite{bass,bord}. Up to now there are no stringent proofs that such approximations
really yield a solution of the full transport problem. The only proof is that numerical
solutions provide the asymptotic limit at equilibrium, which can be deduced from
the Boltzmann equation (the BUU equation without potential and Pauli-blocking factor)
(see details in Ref.\onlinecite{bass}).
We recall that the BUU equation is a generalization of
the Fokker-Planck equation associated with the Brownian motion
of particles in a classical Coulomb gas model. Dyson discovered that this model
can be described well in terms of the Gaussian ensembles of
the Random Matrix Theory (RMT) \cite{meh}. This fact
implies that  one might expect that the RMT could be useful
for analysis of experimental results
obtained in heavy ion reactions at high energies.

The RMT  was originally introduced
to explain the statistical fluctuations of neutron resonances in compound
nuclei \cite{P65} (see also Ref.\ \onlinecite{Brody}).  There the precise
heavy-compound-nuclear Hamiltonian is unknown or rather poorly known, and there
is a large number of strongly interacting degrees of freedom.  Wigner first suggested
replacing it by an ensemble of Hamiltonians which describe {\it all possible
interactions} \cite{Wigner}.  The theory assumes that the Hamiltonian belongs to
an ensemble of random matrices that are consistent with the fundamental symmetries
of the system.  In particular, since the nuclear interaction preserves time-reversal
symmetry, the relevant ensemble is the Gaussian Orthogonal Ensemble (GOE).  Whereas
if time-reversal symmetry were broken, the Gaussian Unitary Ensemble (GUE) would
be the relevant ensemble. The GOE and GUE correspond to ensembles of real symmetric
matrices and of complex Hermitian matrices, respectively.

It is noteworthy that during the last twenty years the RMT  grew into
the powerful new statistical theory of fluctuations in a variety of physical problems.
For example, the RMT has become a very successful theoretical
method for analysing fluctuation properties found in the data from atoms and nuclei,
quantum dots, and many other systems (see, for example, Ref.\onlinecite{Gur}).
In various fields, the
Dyson-Mehta statistical measures are the most often used to quantify a system's
correlations and to determine what information the fluctuations contain. They
can be used with {\it any discrete data} to search for correlations.
These measures do not depend on the background of measurements and
used in the context of RMT give universal forms depending only on the fundamental
symmetries preserved.  Their only requirement is that local mean densities (or
secular behaviors) be understood and their effects be removed.
Furthermore, a change of fluctuation properties of a system under consideration,
induced by a change its symmetry properties, can be detected by the RMT tools
unambiguously (see below). Therefore, these tools
could provide a way of detecting the transition between different
phases (with different fluctuation properties) of strongly interacting matter
produced in high energy nucleus-nucleus collisions.

The main aim of the present paper is to introduce
the Dyson-Mehta statistical measures for analysis
of experimental data from nucleus-nucleus collisions at high energies in order
to detect effects produced by correlations. More specifically, to demonstrate
that  they are very sensitive to spacing correlations present in the nucleus-nucleus
collision data, and serve a useful purpose of independent complementary approach to
the standard tools designed for such an analysis
in high energy physics. Note, in general, the standard tools are based on
different concepts, in contrast to the Dyson-Mehta  measures that are formulated
within  unique, mathematically rigorous theory.
In Section II we review how to detect the
manifestation of correlations with the aid of nearest neighbor spacing
distribution (NND) from the data obtained in light nuclei collisions in
Dubna experiments.  To check the validity of our findings by dint of the NND
in Section III we analyse the same data within the method of effective mass
widely used in the analysis of data from nucleus-nucleus collisions at high energies.
Section IV is devoted to analysis of density-density correlations based on the
Dyson-Mehta  statistical measures and the two-pair correlation function often used in
high energy physics.  The main conclusions are summarized in Section V.
The preliminary results of our analysis were presented in \cite{sh1}.

\section{Nearest-neighbor spacing distribution}
\label{spac}
\subsection{Some experimental details}
To demonstrate the feasibility of using the Dyson-Mehta statistical measures in
the context of high energy nuclear collisions, we make use of the experimental
data that have been obtained with the 2-m propane bubble chamber of
High Energy Physics Laboratory (LHE), JINR~\cite{[8],[9],int3}. The chamber, placed
in a magnetic field of 1.5 T, was exposed to beams of light relativistic carbon nuclei
at the Dubna Synchrophasotron.
Nearly all secondary particles, emitted in a 4$\pi$ total solid angle, were
detected in the chamber.

\begin{figure}[ht]
\includegraphics[height=0.25\textheight,clip]{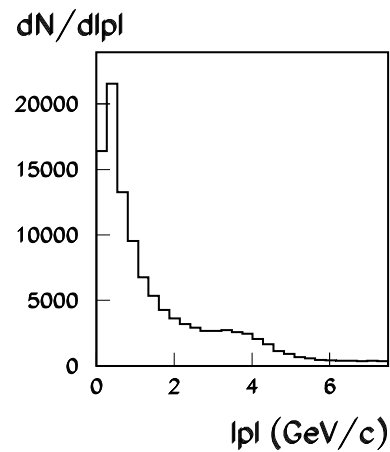}
\caption{
The momentum distribution of the secondary charge particles.
}
\label{mom}
\end{figure}
All negatively charged particles, except those identified as
electrons, are considered as $\pi^-$-mesons. The contamination from
misidentified electrons and negative strange particles does not exceed 5$\%$
and 1$\%$, respectively.
On Fig.\ref{mom} the momentum distribution $dN/d|p|$
of the secondary charged particles is displayed.
The maximum of the particle production is observed at $|p|\sim 0.5 - 0.7$ GeV/c.
The particle momenta were calculated from
the particle trajectories in a magnetic field taking into account ionization and
radiation losses. In particular, the uncertainty in the momentum value were estimated
taking account of  the effects of multiple Coulomb scattering and bremsstrahlung radiation.
The average uncertainty in the momentum and the angle measurements varies as:
$\langle \Delta p/p\rangle=(11.5\pm 0.3)\%$, $\langle \Delta \theta \rangle \sim 0.8^0$.
The minimum momentum for pion registration is about
70 MeV/c in lab frame (we consider all kinematic variables in the lab frame).
The protons were selected by a statistical method applied to all positive
particles with a momentum of $|p|>500$ MeV/c (slow protons
with $|p|\le 700$ MeV/c were identified by ionization in the chamber).

In this experiment,
there are 37792 ${}^{12}$C+${}^{12}$C interaction events at a momentum of 4.2A GeV/c
(for greater discussion of the details see~\cite{[9],int3,[9a]}) containing
7740 events with more than ten tracks of charged particles.
The separation method of ${}^{12}$C+${}^{12}$C collisions in propane ($C_3H_8$), data
processing, identification of particles and discussion of corrections are described in
detail in \cite{[10]}. About 70$\%$ of ${}^{12}$C+${}^{12}$C events have been
identified with the aid of this method.
The set residue of events was separated
statistically between carbon+carbon and carbon+hydrogen collisions with the aid
of the event "weight". The "weights" were estimated from known cross sections for
inelastic carbon+carbon and carbon+hydrogen collisions.

\subsection{Basic remarks}
\label{basic}
If one supposes that the momenta of secondary particles produced
in nucleus-nucleus collisions may be treated in analogy with eigenstates of
a composite system, just like the eigenstates of the compound nucleus, then
the statistical analysis methods \cite{P65} introduced by Dyson and Mehta can
be applied to the LHE collision data.
By analogy with compound nuclei discussed in the Introduction, one may assume
that the observed spectra (Fig.\ref{mom}) belong to the effective QCD Hamiltonian
which describes the composite system, but is unknown.
We point out that the secondary particle momenta
are well determined in the experiment. It is, therefore, natural to use
the momentum as a proper variable for our analysis
in order to avoid possible errors
at the transformation from the momentum to the energy, which requires
an accurate determination of the corresponding mass value.

\begin{figure*}[ht]
\includegraphics[height=0.5\textheight,clip]{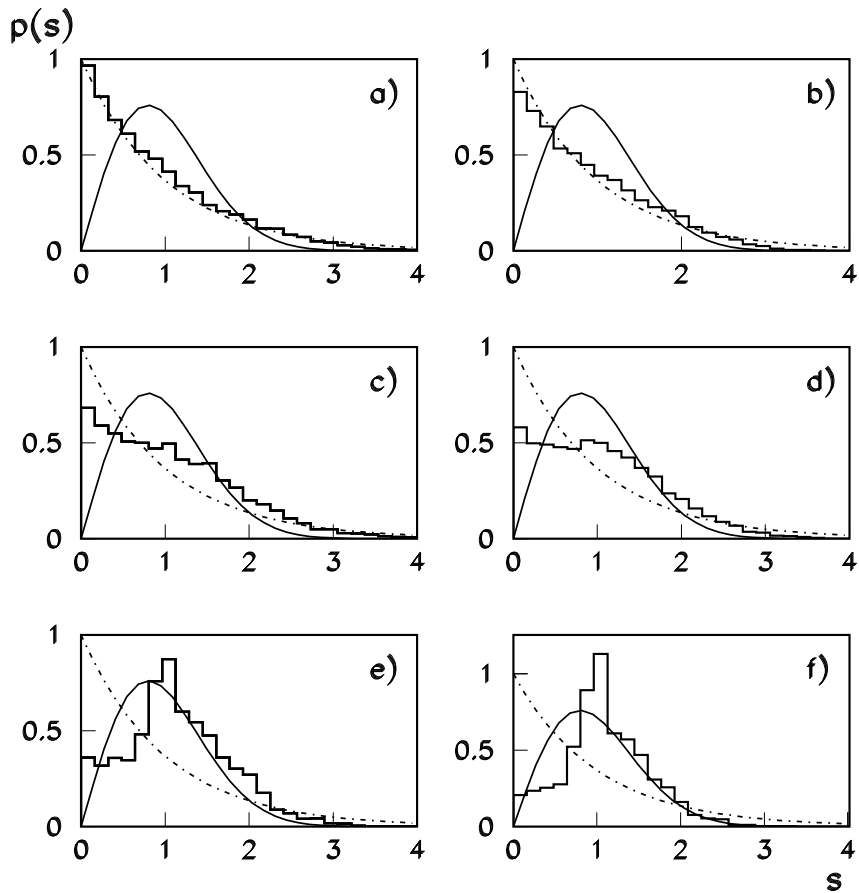}
\caption{
Nearest-neighbor spacing momentum distribution (histogram)
p(s) for different regions of measured momenta: (a) and (b)
correspond to $0.15 < |p| < 1.14$ GeV/c;
(c) and (d) correspond to $1.14 < |p| < 4.0$ GeV/c;
(e) and (f) correspond to $4.0 < |p| < 7.5$ GeV/c.
The total distribution (it includes data related to all charged secondary particles)
consists in (a), (c), and (e) panels.
The distribution for proton momenta consists in
(b), (d), and (f) panels.
The Poisson and the Wigner surmise distributions are connected
by dot-dashed and solid lines, respectively.}
\label{fig1}
\end{figure*}

 Based on this supposition, the ordered sequence
of ``energy levels'' $\{E_i\}, i=1,...,N$ comes from
the ordered sequence of the detected momenta of the secondary particles.
Hereafter, for the sake of convenience we associate a symbol "E" with
an absolute value of the momentum $|p|$.
To analyse the statistical fluctuations of the spectrum $\{E_i\}$, it is necessary
to separate its smoothed average part whose behaviour is nonuniversal and cannot
be described by the RMT \cite{boh2}. A separation of fluctuations of a quantum spectrum
can be based on the analysis of the density of states below some threshold $E$
\beq
\label{d}
\rho(E)=\sum_{i=1}^N\delta(E-E_i) \, .
\eeq
We can define a staircase function
\beq
\label{s}
S(E)=\int_{-\infty}^E\rho(E')dE'=\sum_{i=1}^N\theta (E-E'),
\eeq
giving the number of points on the momentum axis which are below
or equal to $E($or $|p|)$.
Here
\beq
\label{t}
\theta (x) \,=\,
\left \{
\begin{array}{l}
0\qquad \,for \,\,\, x < 0 \\
1\qquad \,for \,\,\, x > 1 \\
\end{array}
\right.
\eeq
We separate $S(E)$ in a smooth part $\zeta(E)$ and the reminder that will
define the fluctuating part $S_{\rm fl}(E)$
\beq
\label{s1}
S(E)=\zeta(E)+S_{\rm fl}(E),
\eeq
such that the integral of $S_{\rm fl}(E)$ is zero.
Here, the smooth function $\zeta(E)$ gives the mean number of eigenvalues less
than or equal to $E$ of the exact eigenvalue distribution $S(E)$.
The smooth part $\zeta(E)$ can be determined either from semiclassical arguments
or by using a polynomial/spline interpolation for the exact staircase function.
From this sequence a new one is obtained  by the unfolding procedure
of the original spectrum $\{E_i\}$ through the mapping $E \rightarrow x$
(see details in Sec.\ref{exhis})
\beq
\label{m}
x_i=\zeta(E_i), \qquad i=1,...,N\,.
\eeq
The effect of the mapping is that the sequence $\{x_i\}$ has on average a
constant mean spacing (or a constant density), irrespective of the
particular form of the function $\zeta(E)$ \cite{boh2}.
Note that there is no any combinatorics involved in a such procedure.
That which remains in the sequence are the fluctuations away from unit mean.
Next, one defines the spacing $s_i=x_{i+1}-x_i$ between two adjacent points. Collecting
${s_i}$ in a histogram, one obtains the probability density or the NND.
We stress that, in general, this procedure does not involve
any uncertainty or spurious contributions and deals with a direct processing of
physical data.

Bohigas {\it et al} \cite{Boh} conjectured that the RMT
describes the statistical fluctuations of a quantum system whose classical
dynamics is chaotic. Quantum spectra of such systems manifest a strong
repulsion (anticrossing) between quantum levels, while in non-chaotic
(regular) systems crossings are a dominant feature of spectra. In turn,
the crossings are usually observed
where there is no mixing between states that are characterized by different good
quantum numbers, and the anticrossings signal a strong mixing due to
a perturbation brought about by either external or internal sources
(cf Refs.\onlinecite{P65,Brody}).

Thus, if the "events" $\{x_i\}$
are independent, i.e., correlations in the system under consideration
are absent, the form of the histogram must follow $p(s)=\exp(-s)$ known
as the Poisson density.  On the other hand, if the levels are repelled
(anticorrelated) as in the GOE, the density is approximately given by the Wigner
surmise form $p(s)=\frac{\pi}{2}s \exp(-\frac{\pi}{4}s^2)$.
In other words, any correlations that produces the deviation from
the regular pattern (Poisson distribution): production of a collective state
(resonance), or some structural changes in the system under consideration
would be uniquely identified from the change of the histogram shape.
In particular, the transition from one probability density to the other has
been used in nuclear structure physics to study the stabilization of
octupole deformed shapes and transition from the chaotic to the
regular pattern in the classical limit \cite{H94}.   Therefore, such an analysis
can provide the first hint of some structural changes at different parameters of
the system under consideration, in particular, in
different energy (momentum) intervals.

\subsection{The experimental NND histograms}
\label{exhis}

In order to reveal structural changes in the momentum distribution
we determine the smooth part $\zeta(E)$
using a polynomial function of the fifth order to interpolate
the exact staircase function (see also Eq.(\ref{s1})):
\begin{equation}
\label{fex}
\zeta(E)=\sum_{k=0}^{5}a_k E^k,
\end{equation}
where $E=|p|$ are the experimental momentum values in
the given event. We recall that momenta are ordered in ascending series.
The parameters $a_k$ were optimized with the aid of the program MINUIT.
Next, we obtain a new spectra $\{x_i\}$ by the unfolding procedure
of the original spectrum $\{E_i\}$ through the mapping $E \rightarrow x$
by means of the polynomial function (\ref{fex})
with the optimized coefficients $a_k$: $x_i=\zeta(E_i), i=1,...,N$.
Having a set of variables $\{x_i\}$, one is  able to determine
the set of spacings $\{s_i\}$.
Since all events are independent we apply the same procedure
for the other events and obtain the new independent
sets of spacings $\{s_i\}$.
To reduce the uncertainty in the data
we put an additional constraint: the distributions of various
spacings $s_i$ from the 7740 events satisfy the condition of $\chi^2$ per degree
of freedom is less than unity (in our case six parameters).
Collecting all independent spacing
together we require the fulfillment of the following conditions:
\begin{equation}
   \int_{0}^\infty p(s)ds =1,\quad
   \int_{0}^\infty s p(s)ds =1.
\end{equation}
The first condition provides the normalization of the probability, while
the second one gives the normalization of the mean distance to unity.

To clearly recognize correlations we
divided the total set of spacings $\{s_i\}$ on three sets, in correspondence
with three regions of   the measured momenta: a)$0.1<|p|<1.14$ GeV/c (region I);
b)$1.14<|p|<4.0$ GeV/c (region II); c)$4.0<|p|<7.5$ GeV/c (region III)
(see Fig.\ref{fig1}).
The region boundaries were determined with the requirement that
the shape of the spacing density  $p(s)$ does not change in the region
under consideration. For example, the decrease of the upper bound of the region I
does not change the Poisson distribution (see below).
However, the increase of this boundary leads to the deviation from the latter
distribution.  The same criteria has been applied for the other two regions.
Note that there is no a prescribed procedure how to define such regions.
However, the empirical approach described above proved to be useful in
data processing for various systems at the RMT analysis (see references in
Refs.\onlinecite{Brody,Gur}). We also separated the distributions associated
with the proton momenta in order to illuminate possible effects related to
the mass difference between proton and pions.
Although there is the distinction between the total and proton NNDs,
a general pattern is similar in the both cases
(compare left and right columns of Fig.\ref{fig1}).
Thus, the mass difference does not change  the
overall conclusions and it is important for the utility of the
approach.

The NNDs in the region 0.15-1.14 GeV/c
are displayed on Figs.1a,b, where the minimum value of the proton
momentum is 0.15 GeV/c. One observes an uncorrelated pattern:
the function $p(s)$ has the Poisson distribution. In this region
the momentum distribution was defined with a high accuracy.
The region II covers the values 1.14-4.0 GeV/c (Figs.1c,d).
This region  corresponds
to the intermediate situation, where the spacing distribution
lies between the Poisson and the Wigner
distributions. One may conclude that in this region there is already the onset
of some correlations.  These correlations manifest themselves much stronger in the
region 4.0-7.5 GeV/c (region III, see Figs.1e,f). The region III is characterized
by the Wigner-type distribution for the spacing probability.

\begin{figure}[ht]
\includegraphics[height=0.23\textheight,clip]{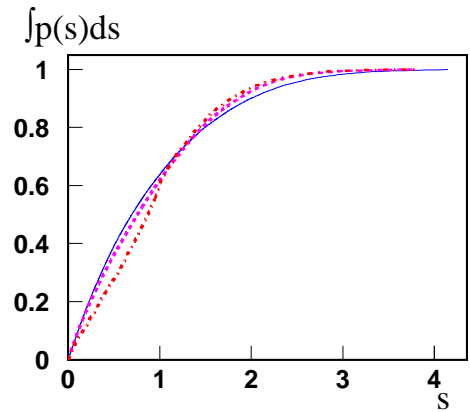}
\caption{(Color online) Cumulative  spacing momentum distributions
            for different  regions of measured momenta:
        a)$0.1<|p|<1.14$ GeV/c (solid line); b)$1.14<|p|<4.0$GeV/c (dashed line);
        c)$4.0<|p|<7.5$ GeV/c (dot-dashed line).
        The solid line nearly coincides with the result for  the Poisson distribution,
        while the dot-dashed line is close to the Wigner surmise distribution.}
\label{fig2}
\end{figure}

From the above analysis one may conclude that in this particular experiment
the method enables us to reveal structural changes in the system at particular
energy regions.
At high energies (the third interval) the particles are produced at
relatively small spatial separation, since
$\Delta p$ is large and one can make use of the uncertainty relation
$\Delta p \Delta x\sim\hbar$.
Thus, one might expect enhanced  correlations.
These correlations continue to persist
at the second  interval but with a lesser intensity. And, finally, in
the first interval the secondary particles move independently in the spatial separation
defined by the kinematics.
To illuminate the physical nature of this observation we consider below the
method of effective mass, traditionally used to process the data produced
at high energy nucleus-nucleus collisions.

To complete the discussion we calculate  the integrated
momentum spacing density for experimental data in order to confirm that all
distributions have on average a constant mean spacing ($= 1$).
This is an additional test which verifies the convergence of our
calculations.
Fig.\ref{fig2} shows the integrated density, demonstrating a
gradual transition from a Poisson-like density toward
a Wigner-like density with the increase of
the absolute value of the momentum distributions.

\section{Method of effective mass}
\label{phys}
In this section we consider
the method of effective mass (MEM) \cite{b4}, which is a standard tool
to extract the information on the correlation between secondary particles.

Before we proceed a few remarks are in order:  in various theoretical approaches
it is usually assumed that at high-energy nucleus-nucleus collisions:
i) a majority of produced secondary pions are emitted basically
through the mechanism of production and decay of the light resonances;
ii) a significant portion of the protons are produced as a result of $\Delta$-isobar
decays \cite{b9}.  Further, it is also assumed that such processes consist
of two steps. First, $(n-k)$ particles and the resonance are produced.
Second, the resonance decays on $k$-particles and one may
expect that due to kinematics there are some correlations between these $k$-particles.
In order to extract these correlations we are forced to consider all possible
combinations of the $k$-particle-participants and compare with
the known resonance masses.

In particular, in the MEM one considers the effective mass for k secondary
particles which create a cluster (a resonance) as
\begin{equation}
M_{k} =\biggl\{(\sum_{i=1}^k E_i)^2-(\sum_{i=1}^k{\vec p}_i)^2\biggr\}^{1/2}\,.
\label{efm}
\end{equation}
If the condition $M_{k} =M_{k}^{res}$ holds, one may conclude that
a resonance with the mass $M_{k}^{res}$ is identified
in the data.  In addition, one assumes that each identified resonance contributes
to the total cross section with its own weight (probability). To carry on this idea,
each resonance is approximated by the Breit-Wigner distribution with
the identified mass and the resonance width. Varying the weights of  identified
resonances, one is aiming to reach the best agreement with the observed
inclusive cross sections of resonances production.
This procedure proved to be reasonable at
small multiplicity of the secondary particles or at relatively small energies per
particle ($< 10$ GeV/c). The situation changes drastically
at the large multiplicity, where number of the secondary particles is more than ten
in each event.
For example, suppose that we have $N$ secondary particles in each
event.  The most favorable case is where each pair of
the secondary particles creates a resonance; the number of resonances
$N_{\it res}=N/2$ in the event. However, we have to consider all independent
combinations between pairs in the event and, therefore, the total
number of physical and "spurious" resonances is $N_{\it tot}=N*(N-1)/2$. It results
in the ratio $r=N_{\it res}/N_{\it tot} \sim 0.1$ (a "useful" signal)
for ten particles in the event. Evidently, this procedure gives rise to
a large spurious  contribution to the analysis.

\begin{figure}[ht]
\includegraphics[height=0.35\textheight,clip]{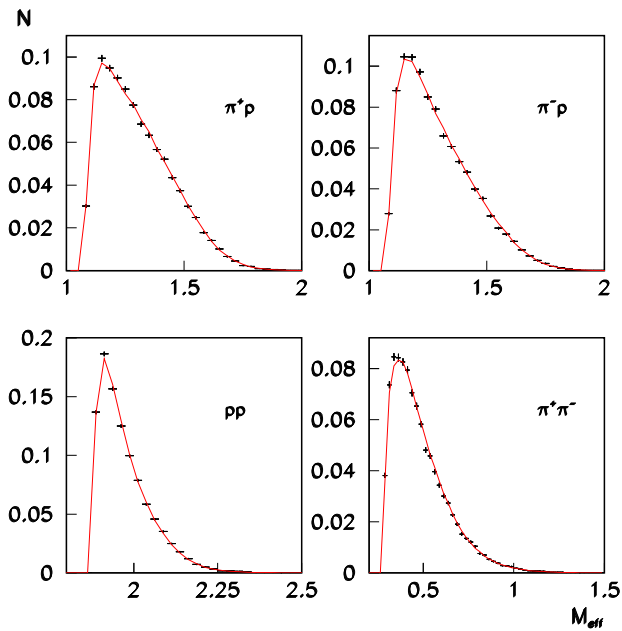}
\caption{(Color online) The distributions (crosses) of
$(\pi^+p)$--, $(\pi^-p)$--, $(pp)$--, and $(\pi^+\pi^-)$--pairs
emitted in the interval of the momentum distribution of the secondary particles
$0.1<|p|<1.14$ GeV/c. The background is shown by solid line. The total number of
pairs is normalized to unity.}
\label{fig3}
\end{figure}

Further, to extract the information on
the resonance production from data
it is necessary to evaluate a background contribution. It is indeed a difficult
task which cannot be solved completely.
Therefore, to construct the background
the method of mixing events \cite{mme} is usually used.
In this method the background is defined by the pairs constructed from the
particles produced in different events only. Note, that this procedure
may lead to the violation of the energy and momentum law conservation due
to the mixing of {\it different} events. Finally, the resonance production
is associated with the existence of the difference between the distribution of
all independent combinations ($N_{\it tot}=N*(N-1)/2$) in
the event and the background distribution for the pairs.

\begin{figure}[ht]
\includegraphics[height=0.35\textheight,clip]{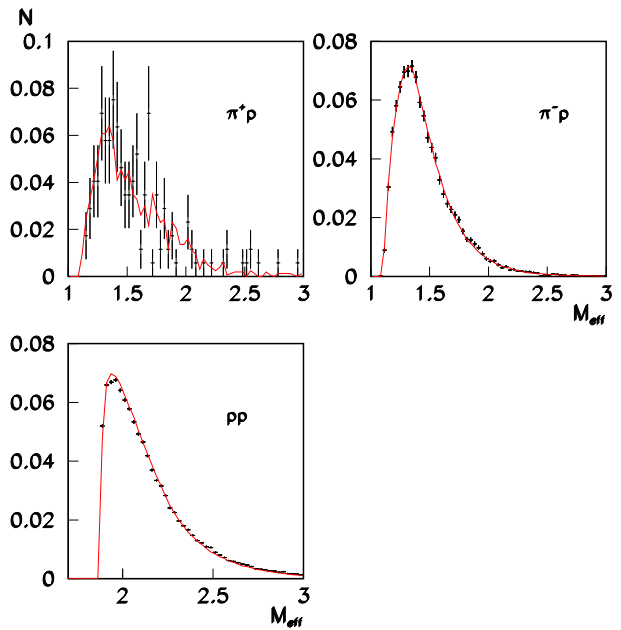}
\caption{(Color online) Similar to Fig.\ref{fig3} for the interval
$1.14-4.0$ GeV/c.}
\label{fig4}
\end{figure}

\begin{figure}[ht]
\includegraphics[height=0.26\textheight,clip]{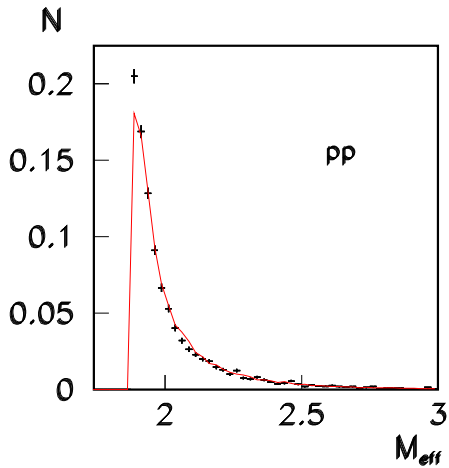}
\caption{(Color online)  Similar to Fig.\ref{fig3}  for
$(pp)$--pairs emitted in the interval of the momentum distribution
of the secondary particles $4.0-7.5$ GeV/c.
}
\label{fig5}
\end{figure}

Keeping this in mind, we consider the distributions of charged particle pairs
emitted in ${}^{12}$C+${}^{12}$C -interactions at $4.2$ A GeV/c \cite{int3,[9]}.
Figs.\ref{fig3}-\ref{fig5} demonstrate the distributions
of $(\pi^+p)$--, $(\pi^-p)$--, $(pp)$--, and $(\pi^+\pi^-)$--pairs emitted in
three ranges of the momentum distribution of the secondary particles:
$0.1-1.14$ (Fig.\ref{fig3}); $1.14-4.0$ (Fig.\ref{fig4})
and $4.0-7.5$ GeV/c (Fig.\ref{fig5}).
All distributions are normalized to the total number of pairs.
Note, the lowest threshold for the effective masses scale is determined by
the individual mass of each participant  and their momenta,
according to Eq.(\ref{efm}).

One observes that in the interval of momentum $0.1-1.14$  GeV/c (Fig.\ref{fig3})
no clear-cut distinction exists between experimental and background distributions.
In this interval we have obtained very good statistical conditions:
41615 $(\pi^+p)$--, 43626 $(\pi^-p)$--, 52992 $(pp)$--
and  39112 $(\pi^+\pi^-)$--pairs. One concludes that there is no
a manifestation of a resonance production. We recall that in this
interval the RMT approach produces the Poisson distribution for the behavior of
the $p(s)$-function (see Fig.\ref{fig1}a). In other words, the NND indicates on the
absence of any correlations in this energy interval.

In the second interval (see Fig.\ref{fig4}) we have
obtained  173 $(\pi^+p)$--, 16470 $(\pi^-p)$--, 167094 $(pp)$--
and  only 16 $(\pi^+\pi^-)$--pairs (are not shown).
There is some deviation of the experimental distribution from the background only
for  $(\pi^+p)$--pairs. Although the error bars are noticeable,
there is a clear indication on the resonance production at
$1.1<M_{\it eff}<2$ (GeV$/c^2$).
We believe that it is connected with the production of
the $\Delta^{++}$-isobars with
masses $m_{\Delta^{++}}=1.232$ and $1.650$ (GeV$/c^2$).
Note that the observation of $\Delta^{++}$-isobars with masses
$m_{\Delta^{++}}=1.232$  GeV$/c^2$ has been  reported in \cite{b9},
while the ones with masses $1.650$ GeV$/c^2$ are observed
for the first time in the present data. The statistics are
not enough, however, to support strongly  the latter observation.
There are no visible deviations between the
experimental data and the background for the distribution of
the $(\pi^-p)$-- and  $(pp)$--pairs.
In this interval the RMT approach produces a visible deviation
from the Poisson distribution for the behavior of $p(s)$-function
(see Fig.\ref{fig1}c), i.e., the NND signals
upon the onset of correlations. The agreement between predictions
obtained in two different approaches confirms that the NND
is able to provide a hint of the resonance production.

In the third interval of the momentum distribution
of charged secondary particles we have
10 $(\pi^{-}p)$-- (are not shown) and 9522 $(pp)$--pairs (Fig.\ref{fig5}).
Here,  the $(\pi^{+}p)$--pairs are absent.
One observes a clear cut distinction between the background and
experimental distribution of $(pp)$--pairs in $4.0-7.5$ GeV/c.
The deviation of the signal relative to the background is above $20\%$.
However, there is no a solid basis to associate such a strong deviation
with a production of di-baryon resonances \cite{debar}, since
the inclusive cross section of such "resonances" would exceed essentially
those that are predicted by various theoretical models.

It is well known that in this interval
the stripping protons are the dominant ones
(with a small contribution of deuterons, tritons and  others) \cite{int3}.
These protons carry a maximum momentum near the value 4.2 GeV/c. It results in
very small  deviations of the particle trajectories in the magnetic field of
the setup. In fact, it is the worst situation for
the accurate determination of the errors in  the momentum distribution.
The RMT approach produces in this interval a distribution of the density
$p(s)$ close to the Wigner surmise form
(see Fig.\ref{fig1}e). As stressed above, such a distribution
is associated with the breaking of regularity in the spectral properties
of a quantum system due to either external or internal sources.
In \cite{sh1} we have already mentioned that the onset of the
Wigner distribution for the density $p(s)$ (breaking of the regularity)
could indicate the presence of  errors in the measurements, i.e.,
the correlations introduced by  external perturbations.

\begin{figure}[ht]
\includegraphics[height=0.35\textheight,clip]{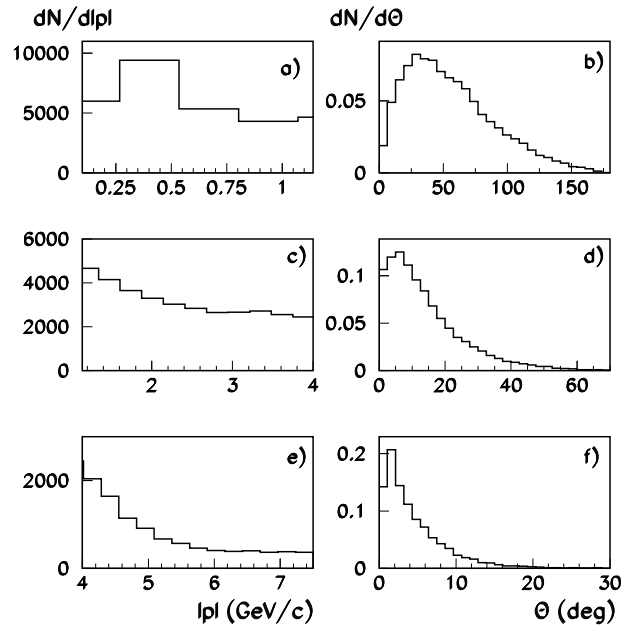}
\caption{
Left: The  proton distributions $dN/d|p|$
as a function of the measured momentum  in three different regions:
$0.1-1.14$ (a); $1.14-4.0$ (c) and $4.0-7.5$ (e) (GeV/c).
Right:  The proton distributions $dN/d\theta$ as a function of the angle in
three different regions: $0.1-1.14$ (b);
 $1.14-4.0$ (d) and $4.0-7.5$ (f) (GeV/c).
The angular distribution is normalized to unity.}
\label{fig6}
\end{figure}

To gain a better insight into the nature of correlations, it is worthy of note that
in all considered distributions there is an evident dominance of
the $pp$-pairs. To trace the evolution of the $pp$ correlations we
select only the momentum and angular distributions
of the protons in three  intervals (see Fig.\ref{fig6}).
In the first interval  the angular distribution (see Fig.\ref{fig6}b) of the pairs
covers almost all angles of the semisphere with some concentration
around $\sim 50^0$. We recall that the NND (see Fig.\ref{fig1}b) displays
the Poisson distribution in this region, which is typical for a noninteracting system.
The absence of the correlations for lower-energy proton pairs is due to Fermi protons
(with the average momentum $\sim 300 $ MeV/c) which move as independent particles.
The angular distribution is determined by
the kinematics (the momentum and the energy conservation laws).
In the second interval the momentum distribution of
the pairs (Fig.\ref{fig6}c) is spread smoothly over all considered values
of the momentum ($1.14-4.0$ GeV/c). There is some concentration of the emitted pairs
in the angular distribution, which covers a solid angle $\sim 20^0$ (Fig.\ref{fig6}d).
The protons can be divided on participants and spectators. The participants are
due to  the $\Delta$-isobar and excited nucleon decay, and also consist of  the protons
which acquire momenta in direct nuclear reactions. Although there are some signature of
spectators, they are dominant in the third region.
For the second region  we obtain the NND pattern which is neither the Poisson nor the Wigner
type (Fig.\ref{fig1}d).
In the third interval (Figs.\ref{fig6}e,f) one observes that
stripping protons have similar momenta and almost zero angle in
the distribution. Evidently, under such conditions, one may expect
a large probability for the interaction in the final state, which leads to
the narrow peak appearances in the effective mass spectrum of the proton pairs.
Such interaction effects in a final state are well known for the particle
production and decay process at high energies \cite{fin}.
We recall that for this interval the RMT approach provides the Wigner distribution
(Fig.\ref{fig1}f) for the behavior of the density $p(s)$. In other words, the NND
indicates that in this region there are strong correlations that break the
regularity in the spectra. The maximum of the correlation is observed to increase
with the increasing of the kinetic energy. This may indicate that the energetic
protons create clusters due to strong pairing effects at short time intervals.
We will return to this point in the next section.

The comparison of the NNDs with the MEM analysis
manifests in fact that there are evident correlations between
behavior of the density $p(s)$ in different energy (momentum) intervals
and the appearance of new sources that break the regularity in
the momentum distribution of the charged particles. The NNDs discussed in the
previous section indicate the onset of correlations
(a production of resonances) at the intermediate region unambiguously,
via the deviation of the spacing distribution from the Poisson limit
(see Fig.{\ref{fig1}}).
According to the MEM analysis the stripping protons are responsible for the
strong correlations in the third interval. We stress that the presence of strong
correlations is predicted independently by the NND which exhibits the Wigner
distribution in this region.
Presently, however, we cannot distinguish between
correlations introduced by the physical process and by possible errors in
the measurements. This is one of the main challenges for the RMT approach in
high energy physics.

\section{Correlations}
\label{cor}

In order to characterize the degree of correlations for a stationary spectrum with
unit average spacing Dyson introduced the {\it k-level correlation functions}
\begin{eqnarray}
&R_k(x_1,...,x_k)=\frac{N!}{(N-k)!}\int ...\int P_k(x_1,...,x_N)dx_{k+1}...dx_N\nonumber\\
&1\leq k\leq N,
\label{p1}
\end{eqnarray}
where $P_N(x_1,...,x_N)dx_{1}...dx_N$ gives the probability of having one eigenvalue
at $x_1$, another at $x_2$..., another at $x_N$ each within the interval
$\{x_i, x_i+dx_i\}$. By integrating $P_k(x_1,...,x_N)$ over all variables but one,
in the
limit $N\rightarrow\infty$, one obtains the ensemble averaged density
\begin{equation}
\tilde{\rho}(x_1)=\int ...\int P_k(x_1,...,x_N)dx_{2}...dx_N\,,
\end{equation}
 which is normalized to unity.  From Eq.(\ref{p1}) it follows that
 $R_1(x_1)=N\tilde{\rho}(x_1)$
and $R_k(x_1,...,x_k)dx_1....dx_k$ is the probability, irrespective  of
the index, of finding  one level within of each of the intervals $[x_i,x_i+ds]$.
From the above definition it follows that $R_1(x)=1$.  With the aid of the
definition (\ref{p1}), by integrating $R_{k+1}$ one obtains
\begin{equation}
\int R_{k+1}(x_1,...,x_{k+1})dx_{k+1}=(N-k)R_{k}(x_1,...,x_{k})\;.
\end{equation}
\begin{figure}[ht]
\includegraphics[height=0.25\textheight,clip]{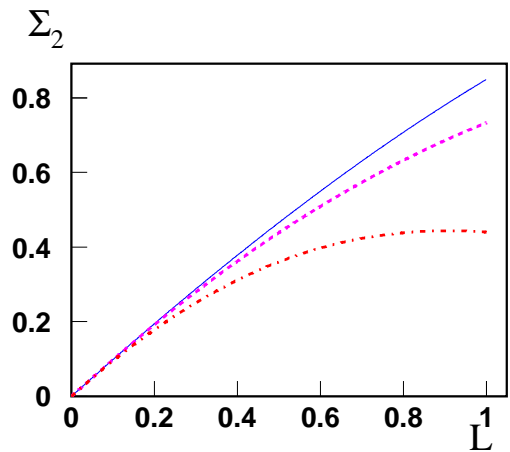}
\caption{(Color online)
Number variance  $\Sigma^2(L)$ for three
different regions of the experimental spacing momentum distributions:
        a)$0.1<|p|<1.14$ GeV/c (solid line); b)$1.14<|p|<4.0$GeV/c (dashed line);
        c)$4.0<|p|<7.5$ GeV/c (dot-dashed line).}
\label{fig7}
\end{figure}

It is difficult to work directly with the $R_k$ functions.  One important and
more convenient measure of correlation that was introduced is based on the number
statistic $n(E,L)$ which is defined to be
the number of levels in an energy interval of length $L$ and
centered at the energy (momentum) $E$:
\begin{equation}
n(E,L)=\int_{E-L/2}^{E+L/2}\rho(E)dE\,.
\end{equation}
Since the spectrum was unfolded (see Eq.(\ref{m})),
the ensemble average number statistic $\langle n(E,L) \rangle =L$ is independent
of the spectrum. However, the variance of $n(E,L)$
\begin{equation}
\Sigma^2(L)=\langle[n(E,L)-\langle n(E,L)\rangle]^2\rangle,
\label{sig}
\end{equation}
does depend on the spectrum considered. For the Poisson density (see \cite{meh})
\begin{equation}
\Sigma^2(L)=L ,
\label{sigP}
\end{equation}
and for the GOE one the exact asymptotic expression is
\begin{eqnarray}
&&\Sigma^2(L)=\frac{2}{\pi^2}\Bigg{[}ln(2\pi L)+\gamma+1+\frac{1}{2}[Si(\pi L)]^2
-\frac{\pi}{2}Si(\pi L)\nonumber\\
&&-\cos(2\pi L)-Ci(2\pi L)+
\pi^2L\left[1-\frac{2}{\pi}Si(2\pi L)\right]\Bigg{]}\,.
\end{eqnarray}
Here $\gamma$ is the Euler constant and $Si$, $Ci$ are the sine and cosine integrals,
respectively.
The number variance $\Sigma^2(L)$ calculated using the optimal implementation of
the definition in Eq.(\ref{sig}) is shown in Fig.\ \ref{fig7}. Indeed, this
quantity manifests the Poisson
statistics (Eq.(\ref{sigP})) for experimental spectra with a low momentum distribution.
On the other hand, one again observes a clear indication of
the presence of correlations for large momenta.

\begin{figure}[hb]
\includegraphics[height=0.22\textheight,clip]{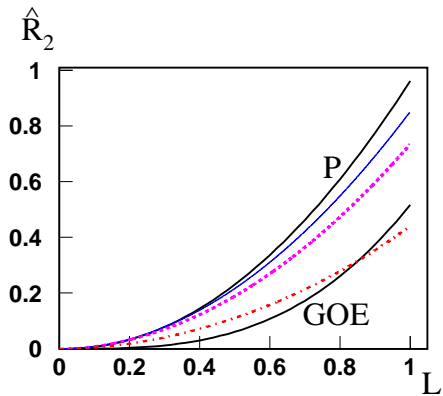}
\caption{(Color online) Two-point  correlation function ${\hat R_2}(s=L)$
                for different  regions of measured momenta:
        a)$0.1<|p|<1.14$ GeV/c (solid line); b)$1.14<|p|<4.0$GeV/c (dashed line);
        c)$4.0<|p|<7.5$ GeV/c (dot-dashed line).
        The solid lines, denoted as P and GOE, display the characteristic
    limits for Poisson and GOE ensembles, respectively.}
\label{fig8}
\end{figure}

For the analysis of fluctuations it is more convenient to use pure
k-point functions \cite{BHP}
\begin{equation}
\hat{R}_k(L)=\int_0^L...\int_0^L R_{k}(x_1,...,x_{k})dx_1...dx_{k}\,.
\end{equation}
The function $\hat{R}_k(L)/k!$ gives the probability that
an interval of length $L$ (for small $L$) contains $k$ levels.
In the RMT most emphasis has been put on the two-point
correlation function ${\hat R_2}(x_1,x_2)$ or density-density correlation function.
The two-point correlation function is the probability density to find two eigenvalues
$x_i$ and $x_j$ at two given energies (momenta), irrespective of the position of all
other eigenvalues. The function $R_2(x_1,x_2)$ depends only on the relative
variables $s=x_1-x_2$ defined in Sec.\ref{spac}.
The variance $\Sigma^2$ is connected to $\hat R_2$ through the following relation
\begin{eqnarray}
\hat R_2(L)=&&\int_0^L\int_0^L R_{2}(x_1,x_{2})dx_1dx_{2}-2\int_0^L(L-s)R_2(s)ds\nonumber\\
=&&\Sigma^2(L)+L(L-1)\,.
\label{r2}
\end{eqnarray}
The two-level correlation function $R_2(x_1,x_2)$  determines
the basic fluctuation measures related to Wigner's level repulsion and
the Dyson-Mehta {\it long-range order}, i.e., large correlations between distant
levels. Bohigas {\it et al} \cite{BHP} provided a thorough analysis
of level repulsion and long-range correlations (rigidity) for different correlation
functions. To understand the distinct role played by level repulsion and long-range order
in the momentum density, we compare our numerical results with
analytical expressions from Table 1 of Ref.\onlinecite{BHP} for the Poisson
ensembles (there is neither level repulsion nor long-range order) and for the
GOE case (this ensemble exhibits both features).  The two-point correlation
function ${\hat R_2}(x_1,x_2)$ calculated from Eq.(\ref{r2}) is shown in Fig.\ref{fig8}.
Even though there are small deviations from the Poisson $(\hat R_2(L)= L^2)$ and
the GOE $(\hat R_2(L)= \pi^2L^3/18)$ predictions, the experimental results for the
momentum distributions reproduce surprisingly well both limits.

We recall that the analysis is based on the unfolded spectra obtained
from the experimental momentum distribution (Sec.II). In contrast to the MEM,
the above procedure does not introduce any "spurious" correlations due to
combinatorics. In addition, each event is treated separately.
However, the two-point correlation function ${\hat R_2}(x_1,x_2)$
clearly exhibits the presence of correlations in the experimental data.
It is another beautiful example how the correlations can show up in the Dyson-Mehta
statistics. Thus, the transition
from the Poisson limit to the GOE implies the onset of strong correlations
at the region III.

As was stressed above the two-point  correlation function ${\hat R_2}(s=L)$
indicates upon large correlations between distant levels (momentum). It appears that the
transition to the GOE limit might signal on the formation of clusters. These clusters should
consist of energetic protons mainly,  which seems
move pairwise in the third interval (see Fig.\ref{fig6}f). Note, however, that
the  NNDs for the third region indicate on the Wigner repulsion.
The physical nature of this phenomenon may be understood with the aid of
the Hanbury-Brown-Twiss (HBT) analysis of the
proton correlations \cite{boal}. The HBT analysis focuses on the spatial distribution
of the radiation area, being very sensitive, however, to  the statistical assumptions
upon the background distribution. Evidently, such analysis represents a separate problem
itself and is beyond the scope of the present paper.

Nevertheless, it is interesting to note, that the standard pair-correlation function
(see, for example, Ref.\onlinecite{gol} and references therein)
\begin{equation}
R(y_1,y_2)=\sigma \frac{d^2\sigma/dy_1dy_2}{(d\sigma/dy_1)(d\sigma/dy_2)}-1,
\end{equation}
used for the analysis of data in high-energy physics, might be useful
to compare with the RMT results.
Here, the quantity $\sigma$ is the cross section of the inclusive reaction and
$y=\frac{1}{2}ln\frac{E+P_{||}}{E-P_{||}}$ is the rapidity, which depends on the
particle energy $E$ and its longitudinal momentum $P_{||}$.
The rapidity is one of the main characteristics widely used in relativistic
nuclear physics (see \cite{b2,b4}). In particular, the change of the reference frame
leads to a trivial shift $\Delta y$ in the rapidity.
The pair-correlation function manifests the difference between the probability density
of two-particle events and the product of the probability densities of independent
particle events. It vanishes if the particle rapidities are independent, i.e.,
{\it the correlations are absent}. In a way, this function is similar in spirit to
the two-point correlation function ${\hat R_2}(x_1,x_2)$, or density-density correlation
function of the RMT.
Therefore, it is useful to compare the predictions obtained by dint of the RMT tools and
with the aid of the standard pair-correlation function.

\begin{figure}[ht]
\includegraphics[height=0.3\textheight,clip]{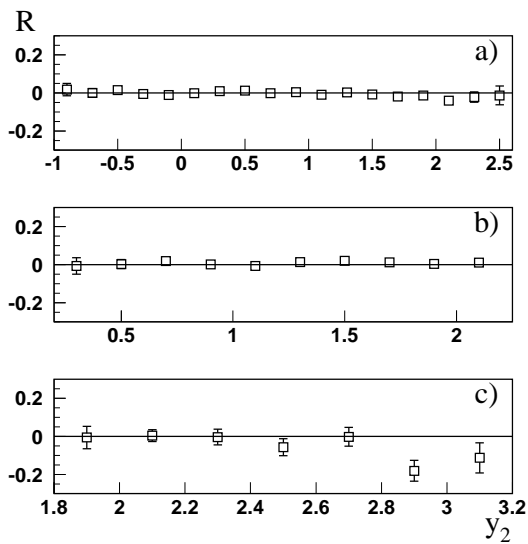}
\caption{Integrated two-pair correlation functions for
        particles obtained in ${}^{12}$C+${}^{12}$C-interactions (see the text)
                for different  regions of measured momenta:
        a)$0.1<|p|<1.14$ GeV/c; b)$1.14<|p|<4.0$GeV/c;
        c)$4.0<|p|<7.5$ GeV/c.}
\label{fig9}
\end{figure}
Fig.\ref{fig9} displays results for the standard pair-correlation function,
obtained for ${}^{12}$C+${}^{12}$C-interactions.  The
function $R(y_1,y_2)$ was integrated over one of the variables, say $y_1$, and we consider
the dependence of this function on $y_2$. For different momentum distributions, there
are three intervals of integration for the variable $y_1$:
a)for $0.1<|p|<1.14$ GeV/c the function $R(y_1,y_2)$ is integrated in the
interval $-1.0<y_1<2.5$;
b)for $1.14<|p|<4.0$ GeV/c  it is integrated in the interval $0.1<y_1<2.4$; and
c)for $4.0<|p|<7.5$ GeV/c it is integrated in the interval $2.5<y_1<3.5$.
The results for the function $R=\int_{y_1}R(y_1,y_2)dy_1$
clearly indicate the presence of correlations between particles
in the region $4.0<|p|<7.5$ GeV/c : there is a strong deviation
from zero in the interval $2.7<y_2<3.2$.
While this function does not exhibit any signature of correlation in the
second region, one observes a strong correlations  between rapidities in the
third region.
We recall that in this interval the NND exhibits
the Wigner distribution (short-range correllations), typical for strongly correlated system.
Thus, we speculate that in the region III there is a formation of
the short-living proton pairs (see also Sec.\ref{phys}).

\section{Summary}
Using the Dyson-Mehta statistical measures for the first time for analysis of data
in high energy physics, we have studied
the experimental data from ${}^{12}$C+${}^{12}$C collisions at 4.2 A GeV/c.
The high accuracy of the data provides a reliable basis for  the
analysis of the momentum distribution of the secondary charged particles
produced in this reaction.
The NND analysis (Sec.II) indicates on the onset of correlations
in the momentum distribution interval $1.14<|p|<4.0$ GeV/c (region II).
The results of calculations for different correlations functions (Secs.II,IV)
evidently demonstrate the presence of strong correlations in the
momentum distribution interval $4.0<|p|<7.5$ GeV/c (region III).

We have analysed the same data with the aid of the method of effective mass,
one of the standard tools for  analysis of data produced at nucleus-nucleus
collisions at high energies (see Sec.\ref{phys}). The results of this analysis
exhibits the presence of strong correlations, independently found by
dint of the RMT approach.
In the region II we found the production of well known $\Delta^{++}$-isobars with
masses $m_{\Delta^{++}}=1.232$ and $1.650$ (GeV$/c^2$).
The dominance of the proton pairs with zero angle in the
momentum distribution interval $4.0<|p|<7.5$ GeV/c
could be attributed to the interaction effects in the final state.
It appears that these effects may lead to the formation of the
short-living proton pairs.
Note that  the method of effective mass involves, however, some additional
assumptions that induce unavoidable uncertainty in the analysis due to a
large nonphysical contribution to the data.
This may cause additional errors, which are difficult to
exclude with the aid of the approaches widely used for the analysis of
nucleus-nucleus collisions at high energies.
In contrast, the RMT approach is independent of such spurious contributions.
The only requirement of the RMT is that the local mean densities (or secular behaviors)
should be understood and their effects removed through the use of the unfolding procedure
(see Sec.\ref{spac}), which is  a routine procedure for the RMT.
By dint of relatively simple and transparent mathematical tools,
the RMT approach detects the appearance of new sources that break the regularity in
the momentum distribution.

Furthermore, the predictions based on the RMT analysis are consistent with
the predictions based on the standard pair-correlation function
$R$ (see Sec.\ref{cor}), which is another tool for the analysis of data
produced at nucleus-nucleus collisions at high energies.
In fact, the RMT two-point correlation function magnifies
the presence of correlations manifested in the standard pair-correlation function
(compare Figs.\ref{fig8} and \ref{fig9}c).
We stress that the uncertainty of the results encountered in
the theoretical analyses of the multiplicity fluctuations in different
statistical ensembles \cite{beg} does not appear in the RMT approach at all.
The Dyson-Mehta statistical measures rely heavily on the fundamental symmetries
preserved in the nucleus-nucleus collisions.

We conclude that the RMT approach is free from various assumptions concerning
the background of the measurements and it provides reliable information about
correlations induced by external or internal perturbations.
We recall that it is expected that at central heavy-ion collisions
at relativistic energies the multiplicity of  secondary particles will
increase drastically. Evidently, the standard tools like the MEM approach
(Sec.\ref{phys})
would encounter a large uncertainty due to the increase of spurious background
contributions in the analysis of data. In contrast, the efficiency of the RMT approach
increases with the increase of the multiplicity of the secondary particles in each event
taken individually. We stress that the RMT approach is based on a
unique, mathematically rigorous theory at the limit of $N\rightarrow \infty$.
Therefore, we believe that our analysis
provides the basis for further development of quantitative methods  based on the
ideas of the RMT for data processing produced at nucleus-nucleus collisions at high
energies.

\section*{Acknowledgments}
We are thankful to High Energy Physics Laboratory of JINR for providing us with
the experimental data and, in particular, to Yu. Panebratsev for an encouraging
support of our studies. We are grateful to  A. Golokhvastov and M. Tokarev
for fruitful discussions.
This work is partly supported
by RFBR Grant No. 08-02-00118 and
by Grant RNP.2.1.1.5409 of the Ministry of Science and
Education of the Russian Federation. S.\ T.\ gratefully acknowledges support
from the US National Science Foundation grant PHY-0555301.

\end{document}